\documentclass{aa}
\usepackage{graphicx}
\usepackage{txfonts}
\begin{document}
 
\title{Pulsation of the Lambda Bootis star HD~210111}
 
\author{M.~Breger\inst{1}, P.~Beck\inst{1}, P.~Lenz\inst{1}, L. Schmitzberger\inst{1},  E. Guggenberger\inst{1},
R. R. Shobbrook\inst{2}}

\offprints{M.~Breger}

\institute{Institut f\"ur Astronomie der Universit\"at Wien, T\"urkenschanzstr. 17,
A--1180 Wien, Austria\\
	\email{michel.breger@univie.ac.at}
\and
Research School of Astronomy and Astrophysics, Australian National University, Canberra, ACT, Australia\\
	\email{rshob@tpg.com.au}}

\date{Accepted 2005 month day.
      Received 2005 month day;
      in original form 2005 month date}

\abstract{The $\lambda$~Bootis stars are a small spectroscopic subgroup of Population I A-type stars and show significant underabundances
of metals. Many are $\delta$~Scuti pulsators.}
{HD~210111 was selected for a detailed multisite pulsation study to determine whether its pulsation properties differ from those
of normal A stars.}{262 hours of high-precision photometry were obtained at the SAAO and SSO observatories.}{13 statistically significant pulsation frequencies were detected with very small photometric amplitudes from 1 to 7 millimag in the visual. A comparison with earlier 1994 measurements indicates a small increase in amplitude. As a byproduct, one of the comparison stars, HD~210571,  was discovered to
be a millimag variable with a frequency of 1.235 cd$^{-1}$ and is probably a new $\gamma$ Doradus variable. The observed wide range of excited frequencies from 12 to 30 cd$^{-1}$ in HD~210111 can be explained with both the
single- and double-star hypothesis. HD~210111 is in a similar evolutionary status to FG Vir, which also shows a wide
range of excited frequencies with a similar frequency spacing near 4 cd$^{-1}$. This is interpreted as successive radial orders
of the excited nonradial modes. In the double-star hypothesis previously evoked for HD~210111, the low and
the high frequencies originate in different stars: here
HD~210111 would resemble $\theta^2$ Tau.}
{The pulsation of the $\lambda$~Bootis star HD~210111 does not
differ from that of normal $\delta$~Scuti stars.}


\titlerunning{HD~210111}

\maketitle
 
\section{Introduction}
The $\lambda$~Bootis stars are chemically peculiar Population I  stars with spectral types from late B to early F.
Spectral analyses of this small group of objects reveal surface underabundances of most Fe-peak elements and
solar abundances of the lighter elements (C, N, O, and S). The extensive review by Paunzen (2004) notes that only a
maximum of about 2\% of all objects in the relevant spectral domain show this peculiarity.
Different theories to explain the spectra of these stars include accretion of interstellar matter, diffusion,
mass-loss or composite spectra of spectroscopic double stars.

Since most $\lambda$~Bootis stars are situated inside the classical instability strip, a large number show $\delta$~Scuti-type pulsation. This
group of pulsators oscillates in a {\em large} variety of radial and especially nonradial modes (for a review see Breger 2000). The diversity of the
pulsation is a consequence of the different stellar properties (such as rotational velocity and duplicity). A detailed study of the pulsations of $\lambda$~Bootis stars might reveal whether the chemical peculiarity is only a surface effect with little expected effect on the pulsation of the whole star.

Paunzen et al. (2002) report that the average pulsation of $\lambda$~Bootis stars differs from that of the average $\delta$~Scuti stars in two ways: incidence and radial order of the pulsation modes. In order to shed light on the $\lambda$~Bootis phenomenon, the study of such differences are very important. They determine that at least 70 \%  of the $\lambda$~Bootis stars inside the instability strip pulsate.
This compares to lower incidence rates of about 1/3 for normal stars found during a large number of different variability surveys.
However, such statistics (even if carefully carried out) need to be compared with great caution, since most pulsators have millimag amplitudes: the detected incidence strongly depends on the observational details: (i) A number of stars not detected to be variable in previous variability surveys have subsequently found to be variable. (ii) Let us consider the star FG~Vir,
for which 75+ modes have been detected already (Breger et al. 2005). Only one mode has a high amplitude (22 mmag in the $y$ passband).
Without this mode, the small amplitudes of the other modes might have resulted in a nondetection during a photometric variability survey.
(iii) The very extensive observational campaigns by the Delta Scuti Network also test the photometric constancy of the multiple comparison stars used. At the millimag level almost all A0-F5 stars are found to be variable (e.g., see HD 210571 discussed below). We strongly suspect that almost all stars inside the lower instability strip are variable. It is, therefore, conceivable that the high incidence of variability of $\lambda$~Bootis stars shows the high photometric quality of the survey undertaken by Paunzen et al.  A detailed examination (including the observational biasses)
of the 'average' amplitude of these multimode pulsators might help solve the question. It is interesting to note that Bohlender et al. (1999) also find
an incidence of nonradial pulsation of more than 50\% in $\lambda$~Bootis stars.

Another difference between the pulsation properties of $\lambda$~Bootis stars and normal $\delta$~Scuti pulsators concerns the
radial overtones in which the stars pulsate. Paunzen et al. demonstrated that the average value of the pulsation constant, $Q$, is
lower for $\lambda$~Bootis stars. An interpretation would be that these stars pulsate in higher radial orders than the normal A stars.

\begin{figure*}
\centering
\includegraphics[bb=22 413 555 748,width = 145mm,clip]{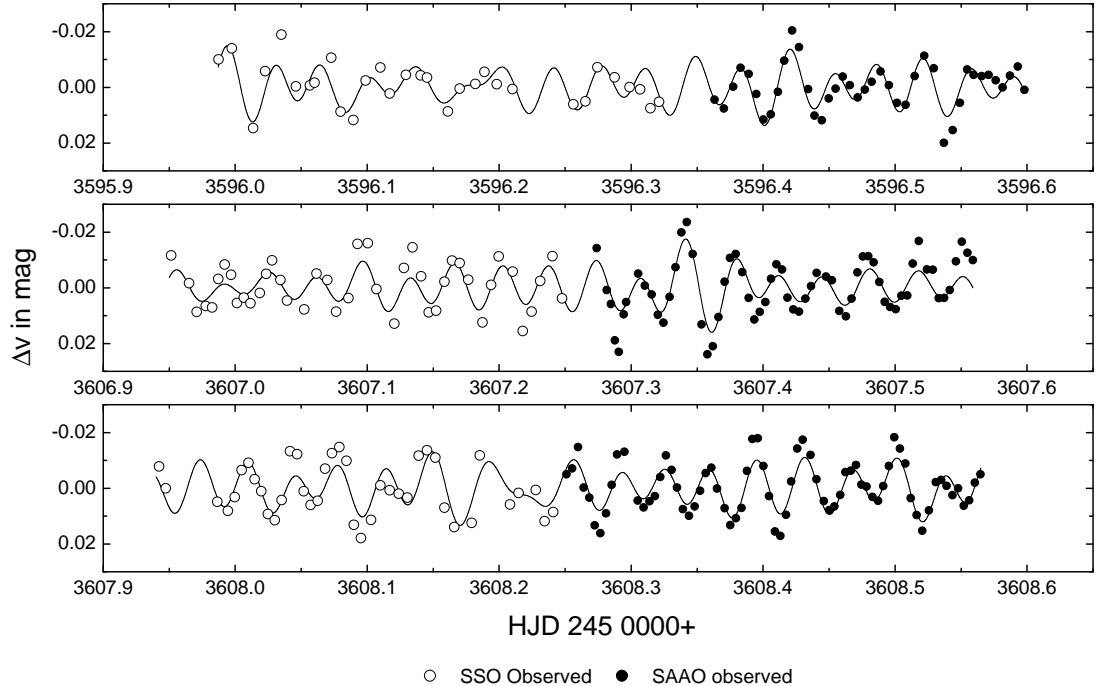}
\caption{Sample light curves of HD~210111. $\Delta v$ are the observed magnitude
differences (variable -- comparison stars) normalized to zero in the narrowband
$uvby$ system. The fit of the multifrequency solution derived in this paper is
shown as a solid curve. The selected light curves were taken from the middle of the campaign
with good coverage, but also show residuals typical for the whole data set.}
\end{figure*}

The Delta Scuti Network utilizes a network of telescopes situated on different continents.
The collaboration reduces the effects of regular daytime observing gaps. The network is engaged
in a long-term program (up to 1000+ hours of observation, photometry and spectroscopy) to
determine the structure and nature of the multiple frequencies of selected $\delta$~Scuti stars
situated in different parts of the classical instability strip.

The $\lambda$~Bootis star HD~210111 (= HR 8437)
was selected for a detailed pulsation study because of its known variability (Paunzen et al. 1994) and
the high value of the published pulsation frequency (28 cd$^{-1}$), suggesting a high radial order. This
high frequency was confirmed by Bohlender et al. (1999), who spectroscopically determined the mode to
be nonradial.

\section{New photometric measurements}

During 2005 July through September, photometric measurements of the star HD~210111
were obtained. Of the scheduled nights, 54 nights with 262 hours were of high photometric
quality. These are listed in Table 1.

The following telescopes were used:

(i) The SAAO measurements were made with the Modular Photometer
attached to the 0.5~m telescope of the South African Astronomical Observatory (SAAO). The observers were
P.~Beck and E. Guggenberger.

(ii) The 0.6-m reflector at Siding Spring Observatory, Australia, was used with a
PMT detector. The observers were P. Lenz, L. Schmitzberger and R. R. Shobbrook.

The measurements were made with Str\"omgren $v$ and $y$ filters. Since telescopes and photometers
at different observatories have different zero-points, the measurements need to be adjusted. This was done by
zeroing the average magnitude of HD~210111 from each site and later readjusting the zero-points by using the
final multifrequency solution. The  shifts were in the submillimag range.

The measurements of HD~210111 (= VAR) were alternated with those of two comparison stars. Details of the
three-star technique can be found in Breger (1993). We used COMP1 = HD~209253 (F6V) and COMP2 = HD~210571 (F3V) as
comparison stars. The spectral type of COMP2 suggests that small-amplitude $\gamma$~Doradus-type
variability cannot be ruled out for this star. In fact, the power spectra of COMP2 relative to COMP1 as well
as VAR showed a clear peak at 1.235 cd$^{-1}$. This is a typical frequency for a $\gamma$~Doradus variable. The
amplitudes were small, viz., 3.8 $\pm$ 0.4 mmag in the $v$ and 2.2 $\pm$ 0.3 mmag in the $y$ passbands. After prewhitening, no other peaks were detected. COMP2 was, therefore, usable after prewhitening. The measurements of VAR were reduced relative to both comparison stars.

A minor problem, which could be solved, concerned small sensitivity drifts of the 0.5~m SAAO photometer. This
eliminated a simple inspection of the Bouguer plots to determine the extinction coefficients. However, the differential extinction
determined from COMP1 and COMP2 agreed with the values determined on the 0.75~m SAAO telescope, which was
used during the same nights.

\begin{table*}
\caption[]{Journal of the PMT observations of HD~210111}
\begin{flushleft}
\begin{tabular}{cclccclc}
\hline
\noalign{\smallskip}
Start & Length & Observatory & Telescope & \hspace{10mm}Start & Length & Observatory & Telescope\\
HJD (days) & (h) & & & \hspace{10mm}HJD (days) & (h) \\
245 0000+ \\
\noalign{\smallskip}
\hline
\noalign{\smallskip}
\noalign{\smallskip}
3562.19	&	3.2	&	SSO	&	0.6 m	&	\hspace{10mm}	3596.36	&	5.6	&	SAAO	&	0.5 m	\\
3568.05	&	1.1	&	SSO	&	0.6 m	&	\hspace{10mm}	3597.16	&	1.2	&	SSO	&	0.6 m	\\
3569.08	&	3.0	&	SSO	&	0.6 m	&	\hspace{10mm}	3597.96	&	4.3	&	SSO	&	0.6 m	\\
3570.07	&	6.4	&	SSO	&	0.6 m	&	\hspace{10mm}	3598.32	&	7.2	&	SAAO	&	0.5 m	\\
3571.10	&	2.0	&	SSO	&	0.6 m	&	\hspace{10mm}	3598.95	&	8.5	&	SSO	&	0.6 m	\\
3572.06	&	1.7	&	SSO	&	0.6 m	&	\hspace{10mm}	3599.31	&	7.5	&	SAAO	&	0.5 m	\\
3573.01	&	6.9	&	SSO	&	0.6 m	&	\hspace{10mm}	3599.94	&	8.7	&	SSO	&	0.6 m	\\
3574.05	&	3.9	&	SSO	&	0.6 m	&	\hspace{10mm}	3601.31	&	6.9	&	SAAO	&	0.5 m	\\
3576.06	&	3.4	&	SSO	&	0.6 m	&	\hspace{10mm}	3602.32	&	7.0	&	SAAO	&	0.5 m	\\
3576.97	&	2.8	&	SSO	&	0.6 m	&	\hspace{10mm}	3604.99	&	7.3	&	SSO	&	0.6 m	\\
3579.11	&	5.4	&	SSO	&	0.6 m	&	\hspace{10mm}	3605.94	&	7.3	&	SSO	&	0.6 m	\\
3580.02	&	4.3	&	SSO	&	0.6 m	&	\hspace{10mm}	3606.95	&	7.1	&	SSO	&	0.6 m	\\
3581.04	&	3.6	&	SSO	&	0.6 m	&	\hspace{10mm}	3607.27	&	6.8	&	SAAO	&	0.5 m	\\
3582.04	&	3.8	&	SSO	&	0.6 m	&	\hspace{10mm}	3607.94	&	7.2	&	SSO	&	0.6 m	\\
3583.05	&	3.6	&	SSO	&	0.6 m	&	\hspace{10mm}	3608.25	&	7.5	&	SAAO	&	0.5 m	\\
3585.04	&	0.3	&	SSO	&	0.6 m	&	\hspace{10mm}	3608.95	&	7.0	&	SSO	&	0.6 m	\\
3589.10	&	5.4	&	SSO	&	0.6 m	&	\hspace{10mm}	3612.28	&	6.8	&	SAAO	&	0.5 m	\\
3590.06	&	1.7	&	SSO	&	0.6 m	&	\hspace{10mm}	3615.15	&	1.6	&	SSO	&	0.6 m	\\
3591.07	&	2.3	&	SSO	&	0.6 m	&	\hspace{10mm}	3615.94	&	2.6	&	SSO	&	0.6 m	\\
3591.97	&	1.2	&	SSO	&	0.6 m	&	\hspace{10mm}	3616.25	&	7.4	&	SAAO	&	0.5 m	\\
3592.47	&	2.5	&	SAAO	&	0.5 m	&	\hspace{10mm}	3617.30	&	6.2	&	SAAO	&	0.5 m	\\
3593.01	&	7.0	&	SSO	&	0.6 m	&	\hspace{10mm}	3617.92	&	0.7	&	SSO	&	0.6 m	\\
3593.36	&	5.8	&	SAAO	&	0.5 m	&	\hspace{10mm}	3618.30	&	2.5	&	SAAO	&	0.5 m	\\
3594.37	&	4.6	&	SAAO	&	0.5 m	&	\hspace{10mm}	3618.92	&	5.8	&	SSO	&	0.6 m	\\
3595.16	&	3.9	&	SSO	&	0.6 m	&	\hspace{10mm}	3619.88	&	8.2	&	SSO	&	0.6 m	\\
3595.35	&	1.9	&	SAAO	&	0.5 m	&	\hspace{10mm}	3627.97	&	5.8	&	SSO	&	0.6 m	\\
3595.99	&	8.0	&	SSO	&	0.6 m	&	\hspace{10mm}	3633.89	&	7.7	&	SSO	&	0.6 m	\\
\noalign{\smallskip}
\hline
\end{tabular}
\newline
\end{flushleft}
\end{table*}

Typical examples of light curves of HD~210111 are shown in Fig. 1 together with the multifrequency solution derived in the next section.

\section{Multiple frequency analysis}

The pulsation frequency analyses were performed with a package of computer
programs with single-frequency and multiple-frequency techniques (PERIOD04,
Lenz \& Breger 2005; http://www.univie.ac.at/tops/period04),
which utilize Fourier as well as multiple-least-squares algorithms. The latter technique fits up to several
hundred simultaneous sinusoidal variations in the magnitude domain and does not rely
on sequential prewhitening. The amplitudes and phases of all modes/frequencies are determined by minimizing
the residuals between the measurements and the fit. The frequencies can also be improved at the same time.

To decrease the noise in the power spectra, we have added the 855 measurements obtained with
the $y$ filter to the 1675 $v$ filter measurements. The dependence of the pulsation amplitude
on wavelength was compensated by multiplying the $y$ data set by an experimentally
determined factor of 1.38 and decreasing the weight of these data points correspondingly.
This scaling creates similar amplitudes but does not falsify the power spectra.
Note that different colors and data sets were only combined to detect new frequency peaks
in the Fourier power spectrum and to determine the significance of the detection.
The effects of imperfect amplitude scaling and small phase shifts between colors
can be shown to be negligible for period finding. For prewhitening, separate solutions were obtained for
each color by multiple least-square fits.

In the analysis of the Delta Scuti Network campaign data, we usually apply
a specific statistical criterion for judging the reality of a newly discovered peak in the
Fourier spectra, viz., a ratio of amplitude signal/noise = 4.0 (see Breger et al. 1993).

Our analysis involves a number of different steps to be repeated. Each step involves the computation
of a Fourier analysis (power spectrum) from the original data or a previously prewhitened fit.
The dominant peaks in the power spectrum were then examined for statistical significance and possible
effects of daily and annual aliasing. For computing new multifrequency solutions, the amplitudes and phases were computed
separately for each color, so that even these small errors associated with combining
different colors were avoided. Note that the new multifrequency solutions were always
computed from the observed (not the
prewhitened) data. Because of the day-time and observing-season (annual) gaps,
different alias possibilities were tried out and the fit with the lowest residuals selected.
The resulting optimum multifrequency solutions were then prewhitened and the analysis repeated
while adding more and more frequencies, until the new peaks were no longer statistically significant.

Fig. 2 shows the results of the frequency search. Altogether 13 statistically significant peaks
in the range of 12.2 to 30.2 cd$^{-1}$ were detected. The solution fits the observations to $\pm$ 3.8 mmag (per
single measurement) in $y$
and $\pm$ 5.1 mmag in $v$. No combination frequencies, f$_i  \pm $f$_j$, were
detected. This can be explained by the small amplitudes of the detected pulsation modes: the
amplitudes of the combination frequencies are expected to be below the detection limit.
The results of the search for multiple frequencies are shown in Table~2.

\begin{figure}
\centering
\includegraphics*[bb=26 43 538 774,width=85mm,clip]{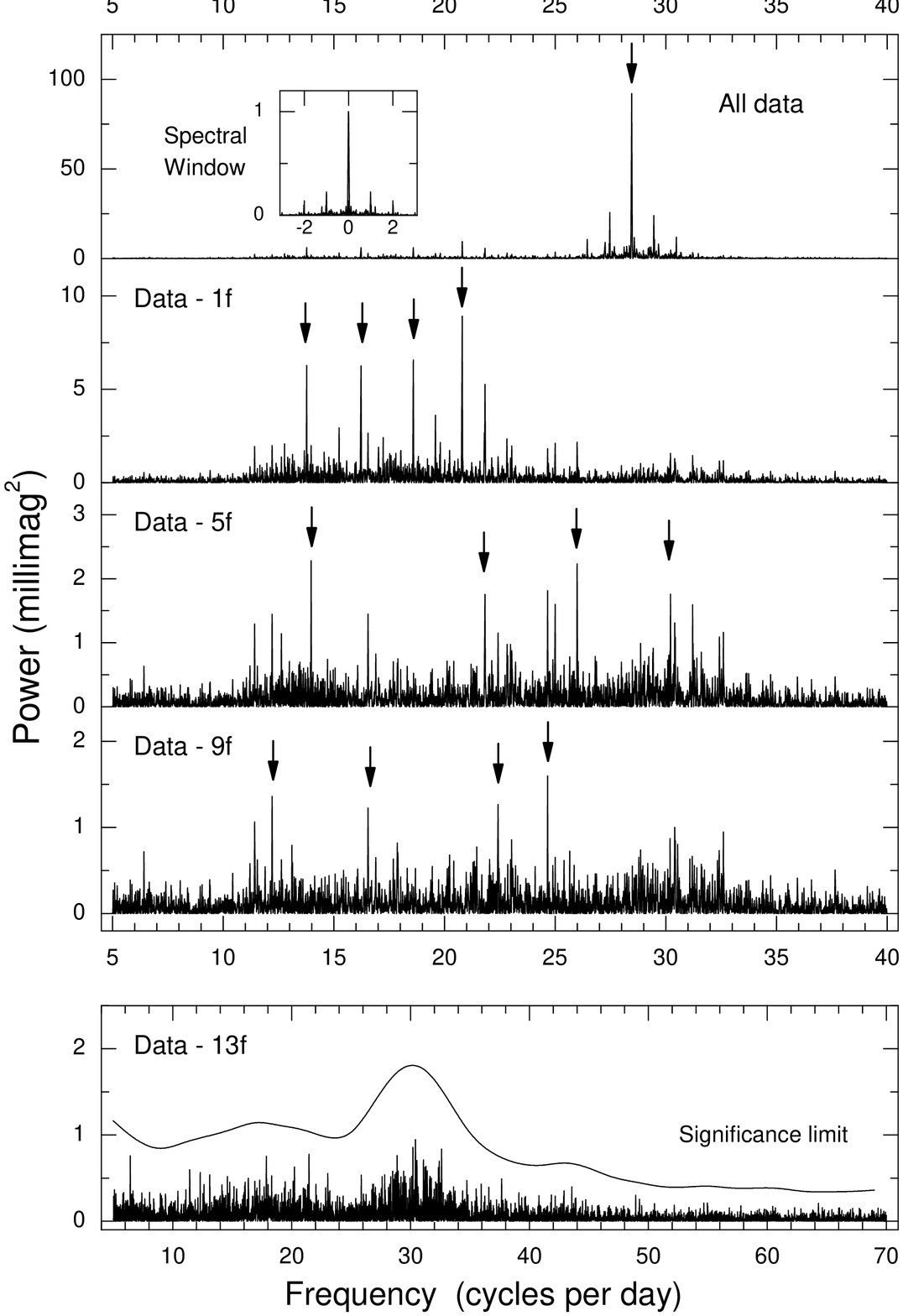}
\caption{Power spectra of HD~210111 for the 2005 photometry.
Top: The dominant mode and the spectral window. Other panels: power spectra after prewhitening
1, 5, 9 and 13-frequency solutions. The bottom panel show that additional modes with frequencies
in the previously detected frequency range are present.}
\end{figure}

\begin{table}
\caption{Detected frequencies of HD~210111}
\begin{center}
\begin{tabular}{lcccc}
\hline
\noalign{\smallskip}
\multicolumn{2}{c}{Frequency} & Detection&  \multicolumn{2}{c}{Amplitude}\\
& & Significance &  $v$ filter & $y$ filter \\
& cd$^{-1}$ & Ampl. S/N & mmag & mmag\\
\noalign{\smallskip}
\hline
& &  & $\pm$ 0.2 & $\pm$ 0.2\\
f$_ {1}$ & 28.47 & 29.0 & 9.5 & 6.9\\
f$_ {2}$ & 20.81 & 10.5 & 2.6 & 1.9\\
f$_ {3}$ & 18.59 & 9.8 & 2.6 & 2.0\\
f$_ {4}$ & 16.23 & 9.3 & 2.4 & 1.9\\
f$_ {5}$ & 13.78 & 9.0 & 2.4 & 1.6\\
f$_ {6}$ & 21.84 & 6.3 & 1.7 & 0.9\\
f$_ {7}$ & 13.98 & 6.2 & 1.4 & 1.4\\
f$_ {8}$ & 24.66 & 6.0 & 1.4 & 1.1\\
f$_ {9}$ & 25.99 & 5.3 & 1.3 & 1.2\\
f$_ {10}$ & 22.43 & 5.1 & 1.2 & 0.9\\
f$_ {11}$ & 12.21 & 4.7 & 1.0 & 1.0\\
f$_ {12}$ & 30.22 & 4.0 &1.2 & 1.3\\
f$_ {13}$ & 16.55 & 4.0 & 1.2 & 0.8\\
\noalign{\smallskip}
\multicolumn{4}{l}{Other interesting peaks in the power spectrum}\\
\noalign{\smallskip}
& 17.87 & 3.5 & 0.9 & 0.6\\
& 6.42 & 3.5 & 0.8 & 0.8\\
& 21.46 & 3.4 & 1.0 & 0.4\\
& 32.60 & 3.2 & 0.9 & 0.7\\
& 30.41 & 3.0 & 1.1 & 0.5\\
\noalign{\smallskip}
\hline
\noalign{\smallskip}
\end{tabular}
\end{center}
A detection is considered significant if the amplitude signal/noise ratio $\geq$ 4.00 (Breger et al. 1993),
This corresponds to a power signal/noise ratio of $\sim$ 12.6. The noise was calculated over 4 cd$^{-1}$ ranges. 

\end{table}

\section{Comparison to previous results}

Paunzen et al. (1994) discovered the variability of HD~210111. The 23 hours of observations obtained in 1994
during 8 nights are sufficient to establish the variability of the star, but insufficient to solve the complex multifrequency
pulsation with very small amplitudes. Consequently, they conclude that the two frequencies (27.99 cd$^{-1}$ and 17.01 cd$^{-1}$)
detected by them can serve only as a guideline for the relevant frequency range.
While the frequency range is confirmed by us, the values of the two frequencies differ from our values by much more than the frequency resolution of the two studies.

We have reanalyzed the 1994 data of Paunzen et al. Their two highest peaks are confirmed by us, but may be
an unfortunate result of severe aliasing of multiple frequencies in a short data set: the 28 cd$^{-1}$ peak
can be shown to be a result of f$_1$ and f$_9$ (28.47 and 25.99 cd$^{-1}$), while the 17 cd$^{-1}$
peak disappears once f$_4$ and f$_7$ (16.23 and 13.98 cd$^{-1}$) are included in the solution. Furthermore, f$_2$ (20.81 cd$^{-1}$) is also present in the 1994 data. Additional conclusions might be an
overinterpretation: however, we note that the formal visual amplitude of f$_1$ in 1994 (3\,$\pm$\,1~mmag)
is smaller than the 7~mmag found in 2005.

\section{Mode identification}

The present measurements were obtained in the $v$ and $y$ passbands in order to
deduce the amplitude ratios and phase lags. These can be a powerful tool for mode
identification (e.g., Daszy\'nska-Daszkiewicz et al. 2005). The application depends on very accurate
determinations of the phase lags, which are difficult to achieve for small amplitudes.
For the main frequency, viz., f$_1$ at 28.47 cd$^{-1}$ we obtain an amplitude ratio A$_v$/A$_y$ = 1.38 $\pm$ 0.05 and
$\phi$(v) - $\phi$(y)= -3.4 $\pm$ 2.0\degr. These values agree with those derived from a smaller sample which includes
only those measurements obtained simultaneously in both passbands (1.36 and -2.9\degr, respectively). 
Alosha A. Pamyatnykh kindly calculated a stellar model with the parameters given in the next section.
The model shows that the observed amplitude ratio and phase shift are consistent with
$\ell$~=~1 pulsation, while the observational uncertainties cannot exclude $\ell$ = 0 or 2.
In this regard, the measured phase shift of HD~210111 is not very helpful for the mode identification of f$_1$.
Nevertheless, our photometric result agrees with the spectroscopically obtained conclusion of
Bohlender et al. (1999) that the mode is nonradial.

\section{Discussion}

The pulsation spectrum of HD~210111 ranges from 12.2 to 30.2 cd$^{-1}$ with the corresponding values of the pulsation constant, $Q$,
ranging from 0.011 to 0.026 d. The amplitudes of pulsation are very small with 7 mmag for the dominant mode and 1 to 2 mmag
for the other 12 frequencies. These small amplitudes are not unusual for $\delta$~Scuti stars: most discovered variables have larger
amplitudes because of observational biasses.

An examination of the spacing between the detected frequencies indicates a preferred frequency
difference around 4.25 and 2.25 cd$^{-1}$. We will now argue that the pulsation properties of HD~210111 are not
unusual and correspond to those found in non-$\lambda$~Bootis stars.

\subsection{Pulsation from a single star}

Let us first assume that the observed pulsation arises in a single star. It is possible to compare HD~210111
with the well-studied normal star FG Vir, for which more than 75 frequencies have been found. Both
stars are similar in temperature and surface gravity:

\begin{flushleft}HD~210111: $T_{\rm eff} = 7550 \pm 100\, $K, $\log \, g = 3.84 \pm 0.15$\\ (Paunzen et al. 2002)

FG Vir: $T_{\rm eff} = 7515 \pm 100\, $K, $\log \, g = 3.99 \pm 0.10$\\
(Breger et al. 1999)
\end{flushleft}

Both stars show pulsation excited over a wide range in frequency of $\sim$20 cd$^{-1}$ or more and have preferred frequency
separations around 4 cd$^{-1}$. Models for this temperature and gravity region show that such
a separation corresponds to successive radial overtones, even among nonradial
modes. The main difference is that for HD~210111 the mode with the highest photometric amplitude (28.47 cd$^{-1}$)  is located near the high-frequency end of the range, while for FG Vir the dominant mode is a low-frequency mode (12.72 cd$^{-1}$). However, there
also exist other $\delta$~Scuti stars with the highest amplitudes at
high frequencies (e.g., 38.1 cd$^{-1}$ in XX Pyx, Handler et al. 2000). The pulsation of HD~210111 is, therefore, not unique or unusual.

\subsection{Pulsation from both components}

Faraggiana et al. (2004) propose that HD~210111 is a double system and mention the possibility that both stars could pulsate. The
observed frequency spectrum would then originate in both stars. Without a clean separation modeling would be impossible.
Stellar systems with multiple pulsators are known, e. g., DG Leo (Lampens et al. 2005) or $\theta^2$ Tau (Breger
2005). For $\theta^2$ Tau a clean separation could be obtained from the study of orbital light-time effects: the frequencies
below 20 cd$^{-1}$ originate in the more evolved object, while the modes near 26 cd$^{-1}$ with very small amplitudes originate
in the main-sequence star. The observed color of HD~210111 does not exclude the possibility of two stars being
inside the pulsation instability strip: the dominant 28.47 cd$^{-1}$ mode would originate in a hot main-sequence component, while
the lower frequencies come from an evolved cooler star. To test this hypothesis, detailed spectroscopic
or very extensive and difficult photometric studies are needed to search for light-time shifts in the different millimag modes.

If the double-star hypothesis is correct, then the pulsation spectrum of HD~210111 would resemble that of $\theta^2$ Tau.

\section{Conclusion}

We have shown that in both the single-star and double-star interpretations, the pulsation of the $\lambda$~
Bootis star HD~210111 does not differ from that of normal $\delta$~Scuti stars. One of the motivations
for the study was the question whether $\lambda$~Bootis stars possess stellar structures different from
those of normal A stars. Such a change would result in a different nonradial pulsation spectrum.
The pulsation results presented in this paper are, therefore, compatible with astrophysical hypotheses that the $\lambda$~Bootis
phenomenon is only a surface effect.

\acknowledgements
It is a pleasure to thank Alosha A. Pamyatnykh for computing a pulsation model for HD~210111 and Ernst Paunzen for helpful comments. This paper uses observations made at the South African Astronomical Observatory (SAAO). This investigation has been supported by the
Austrian Fonds zur F\"{o}rderung der wissenschaftlichen Forschung.


\begin{thebibliography}{}

\bibitem[][{} Bohlender, D. A., Gonzalez, J.-F., \& Matthews, J. M. 1999, A\&A, 350, 1999 
\bibitem[]{} Breger, M. 1993, in Stellar Photometry - Current Techniques and Future Developments, ed. C. J. Butler,
I. Elliott, IAU Coll. 136, 106
\bibitem[]{} Breger, M. 2000, in 'Delta Scuti and Related Stars',  eds. Breger,~M., \& Montgomery, M.~H., ASP Conf. Series, Vol. 210, 3
\bibitem[]{} Breger, M. 2005, in 'The Light-Time Effect in Astrophysics', ed. Sterken, C., ASP Conf. Series, Vol. 335, 85
\bibitem[]{} Breger, M., Stich, J., Garrido, R., et al. 1993, A\&A, 271, 482
\bibitem[]{} Breger, M., Pamyatnykh, A. A., Pikall, H., \& Garrido, R. 1999, A\&A, 341, 151
\bibitem[]{} Breger, M., Lenz, P., Antoci, V., et al. 2005, A\&A, 435, 955
\bibitem[]{} Daszy\'nska-Daszkiewicz, J., Dziembowski, W. A., Pamyatnykh, A. A., et al. 2005, A\&A, 438, 653
\bibitem[]{} Faraggiana, R., Bonifacio, P., Caffau, E., et al. 2004, A\&A 425, 615
\bibitem[]{} Handler, G., Arentoft, T., Shobbrook, R.R., et al. 2000, MNRAS, 318, 511
\bibitem[]{}Lenz, P., \& Breger, M., 2005, CoAst, 146, 53
\bibitem[]{} Lampens, P., Fremat, Y., Garrido, R., et al. 2005, A\&A, 438, 201
\bibitem[]{} Paunzen, E. 2004, in 'The A-Star Puzzle',  eds. Zverko, J., Ziznovsky, J., Adelman, S. J., Weiss, W. W., Cambridge Univ. Press, 443
\bibitem[]{} Paunzen, M., Handler, G., Weiss, W. W., \& North, P. 1994, IBVS, 4094, 1
\bibitem[]{} Paunzen, M., Handler, G., Weiss, W. W., et al. 2002, A\&A, 392, 515
\end{thebibliography}
\end{document}